\documentclass[9pt,twocolumn,twoside]{opticajnl}
\journal{opticajournal} 

\setboolean{shortarticle}{true}

\usepackage{lineno}
\usepackage{mathptmx,graphicx}
\usepackage{siunitx}
\usepackage{soul,color}
\DeclareSIUnit\bar{bar}

\title{Tunable megawatt-scale sub-20~fs visible pulses from a fiber laser source}

\author[1]{Mohammed Sabbah}
\author[2]{Robbie Mears}%
\author[1]{Leah R. Murphy}
\author[2]{Kerrianne Harrington}
\author[2]{James M. Stone}
\author[2]{Tim A. Birks}
\author[1,*]{John C. Travers}

\affil[1]{School of Engineering and Physical Sciences, Heriot-Watt University, Edinburgh,  EH14 4AS, United Kingdom}
\affil[2]{Centre for Photonics and Photonic Materials, Department of Physics, University of Bath, Claverton Down, Bath, BA2 7AY, United Kingdom}

\affil[*]{j.travers@hw.ac.uk}

\begin{abstract}
Ultrafast laser pulses that are both tunable in wavelength and very short in duration are essential tools in fields ranging from biomedical imaging to ultrafast spectroscopy. While resonant dispersive-wave emission in gas-filled hollow-core fibers is a powerful technique for generating such pulses, it has traditionally required complex and expensive pump laser systems. In this work, we present a more compact and accessible alternative that combines gain-managed nonlinear amplification with resonant dispersive-wave emission. Our system produces sub-20~femtosecond pulses tunable from 400~nm to beyond 700~nm, with energies up to 39~nJ and peak powers exceeding 2~MW, operating at a 4.8~MHz repetition rate. This compact and efficient laser source opens new avenues for deploying resonant dispersive-wave-based technologies for broader scientific and industrial applications.
\end{abstract}

\setboolean{displaycopyright}{false} 

\begin{document}
\maketitle

\noindent Tunable ultrafast laser pulse sources with high peak power are widely exploited for applications in the biological and chemical sciences, such as multiphoton imaging and microscopy, and ultrafast spectroscopy~\cite{Denk1990, Maiuri2020, Kotsina_2022}. The most common sources of such pulses in the visible and near-infrared region are tunable Ti:sapphire oscillators, optical parametric amplifiers, or filtered supercontinuum sources. All of these are complex and come with one or more disadvantages in terms of cost, size, complexity, or performance. Although Ti:sapphire oscillators can produce short pulses down to 5~fs~\cite{Ell2001}, scaling to high energy usually produces a significantly longer pulse duration ($\gtrsim\qty{100}{\fs})$. Widely-tunable Ti:sapphire oscillators are the work-horse laser system in multiphoton imaging, but their tunability is naturally limited from $\sim\qty{680}{\nm}$ to $\sim\qty{1080}{\nm}$ without further nonlinear conversion stages that further increase complexity and reduce efficiency. They typically produce $\gtrsim\qty{100}{\fs}$ pulses with $\lesssim\qty{10}{\nano\J}$ energy, resulting in $\sim\qty{100}{\kW}$ peak power.

Alternatively, optical parametric oscillators and amplifiers offer a route to wider tunability, shorter pulses, and higher energy. By incorporating second or third harmonic pump stages, they can be tuned across the visible region~\cite{Cerullo2003, Maiuri2020, Deckert:23, Mevert:21}. For instance, Mevert~et~al.\ demonstrated a high-power visible optical parametric oscillator at 50~MHz producing $\sim 0.45$~W across 480–720~nm; however, the pulse duration was relatively long at $\sim 268$~fs~\cite{Mevert:21}. Few-cycle ultraviolet and visible pulses have been achieved with optical parametric amplifiers~\cite{Cerullo2003}---for example, Deckert~et~al.\ generated 7.8~fs pulses spanning 590--780~nm by pumping a degenerate optical parametric amplifier with the third harmonic of a Yb:KGW laser~\cite{Deckert:23}. Such sources usually require high-energy amplified pump lasers, which makes them complicated and expensive and forces them to operate at a relatively low repetition rate. Filtered supercontinuum sources are a much simpler route to rapid tunability and high average power, but they often have a long pulse duration ($\gtrsim\qty{300}{\fs}$), low pulse energy after filtering, and often lack temporal coherence, prohibiting some applications~\cite{Haohua2013}.

\begin{figure*}
\centering
\includegraphics{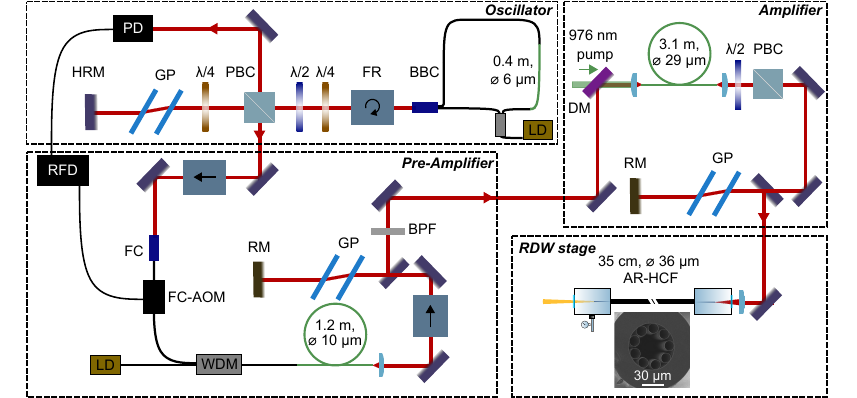}
\caption{\label{fig:setup} Experimental setup. LD: laser diode; WDM; wavelength division multiplexer; BBC: birefringent beam combiner; FR: Faraday rotator; $\lambda/2$, $\lambda/4$: waveplates; PBC: polarizing beam cube; GP: grating pair; HRM: high reflection mirror; PD: photodiode; RFD: radio frequency driver; ISO: isolator; FC: fiber collimator; FC-AOM: fiber-coupled acousto-optic modulator; RM: roof mirror; BPF: bandpass filter; DM: dichroic mirror.  A scanning electron micrograph of the antiresonant hollow-core fiber cross-section is shown in the RDW stage.}
\end{figure*}

Resonant dispersive-wave (RDW) generation in hollow-core fibers has been shown to be an excellent technique for generating tunable ultrashort pulses from the vacuum ultraviolet to the near-infrared \cite{Joly:11, Mak:13, Belli:15, Mak:15, Kottig:17, Travers2019}. The RDW generation process is a result of phase matching between a pump soliton and a linear wave in the presence of high-order dispersion~\cite{Joly:11, Erkintalo:2012}. The duration of the generated RDW pulse can be as short as a few femtoseconds~\cite{Brahms:19, Reduzzi:23}. To avoid modulation instability dynamics and produce high-quality and coherent pulses, the pump soliton order should be $<16$~\cite{dudley_supercontinuum_2006}, which usually requires short pump pulses of $\lesssim\qty{50}{\fs}$, and moderately high energy in gas-filled fibers. By making use of the low guidance loss of small-core antiresonant fibers, pump energy at the few to sub-microjoule level is required. Although this can be further reduced to pump energies as low as 20~nJ after careful optimization in exceedingly small core fibers~\cite{Sabbah:24}, this inherently results in low-energy tunable RDW pulses, thereby limiting their suitability for nonlinear applications that require high peak power. As a result, RDW generation is usually pumped with amplified Ti:sapphire or ytterbium (Yb) laser systems. While Ti:sapphire systems can produce high energy and short pulses (sub-40~fs), sufficient to drive RDW generation directly~\cite{Joly:11, Mak:13, Belli:15}, amplified Yb laser systems produce longer pulses $\gtrsim\qty{200}{\fs}$, requiring an additional temporal compression stage before RDW generation~\cite{Mak:15, Kottig:17}.

Recent advances in fiber lasers and amplifiers have enabled the generation of sub-50~fs pulses with microjoule level energy with a simple architecture~\cite{Liu:19, Repgen:19, Sidorenko:20}. In particular, the recently developed gain-managed nonlinear amplification (GMNA) regime presents a promising approach due to its straightforward design, which eliminates the need for pre-chirping~\cite{Sidorenko:19}. The GMNA approach is essentially an over-extended self-similar nonlinear amplifier~\cite{Fermann2000}, with a longitudinally varying asymmetric gain profile due to gain saturation in the longer fiber. A narrowband, low-energy seed pulse is introduced into the codirectionally pumped gain fiber, where the interaction of spectral broadening with the asymmetric gain dynamics produces a broadband spectrum with an almost linear chirp. The resulting pulse can be compressed close to its transform-limited duration using a simple grating pair~\cite{tomaszewska2022, REN2023, Boulanger:23}.

In this work, we combine a GMNA pump source with RDW emission in gas-filled hollow-core antiresonant fibers to demonstrate an ultrafast pulse source tunable across the near-ultraviolet to the near-infrared, that is compact, efficient, and produces coherent, sub-20~fs pulses with significant energy (up to 39~nJ)---corresponding to more than 2.2~MW peak power---at 4.8~MHz, with a simple architecture.

The experimental setup is shown in Figure~\ref{fig:setup}. It consists of an oscillator, a pre-amplification stage, the GMNA stage, and the RDW generation stage. The oscillator is a self-built all-polarization-maintaining mode-locked Yb fiber laser based on a nonlinear amplifying loop mirror (NALM)~\cite{Jiang2016, Mayer:20}. The NALM oscillator design was chosen for its simplicity, reliable mode-locking, and environmental stability, achieved through the use of polarization-maintaining fibers. Its Sagnac loop-based artificial saturable absorber also offers long-term stability by avoiding the degradation inherent to physical saturable absorbers. The oscillator is modelocked by increasing the laser pump-diode current above the modelocking current threshold. Once modelocking is achieved, the pump-diode current is lowered to maintain a stable and fundamental modelocking state. The oscillator operates at 48~MHz and generates 0.36~nJ pulses. The output pulse duration is 6.7~ps and can be compressed with an external grating pair to 215~fs. The main output of the oscillator is coupled to a fiber-coupled acousto-optic modulator that reduces the repetition rate of the pulse train to 4.8~MHz. The pulses are then amplified in a \qty{10}{\um} core diameter, 1.2~m long Yb gain fiber that is cladding-pumped at 976~nm. The pre-amplifier output pulses are compressed and filtered to 2.2~nm at full-width half-maximum (FWHM), with around 1~ps duration and 0.3~nJ energy.

The GMNA stage consists of a 3.1~m long Yb-doped single-mode polarizing large-mode-area photonic crystal fiber with \qty{31}{\um} mode-field diameter (NKT Photonics). The fiber has a $\sim\qty{10}{\dB\per\m}$ multimode pump absorption at 976~nm. In the GMNA regime, a sub-ps pulse with relatively low energy can be amplified to many hundreds of nJ with sub-40~fs pulse duration~\cite{Sidorenko:19}. The working principle of a GMNA is the interplay between self-phase modulation and the longitudinally varying asymmetric gain profile as the pulse propagates along the fiber, resulting in a broadband pulse with near-perfect monotonic chirp and a few-ps pulse duration, which can be simply compressed to sub-40~fs with a grating pair. After the GMNA fiber, an achromatic half-waveplate combined with a polarizing beam splitter is employed to control the GMNA energy directed to the next stage. The output pulses are subsequently compressed using a 1000~line/mm grating pair.

\begin{figure}[t!]
\centering
\includegraphics{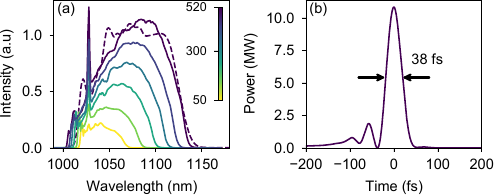}
\caption{\label{fig:GMNA_FROG} (a) The evolution of GMNA output spectrum as the output pulse energy increases; the inset colorbar indicates the corresponding output pulse energy (nJ). (b) Retrieved temporal profile of the compressed GMNA at the maximum output energy. The dashed line in (a) shows the corresponding retrieved spectrum.}
\end{figure}

The RDW generation stage consists of a 35~cm long antiresonant hollow-core fiber with nine resonators, a core diameter of \qty{36}{\um}, and a core-wall thickness of 150~nm. The fiber was fabricated in-house using the usual stack-and-draw technique~\cite{Russell2003,PolettiPetrovichRichardson}. According to the antiresonance model~\cite{Litchinitser:02, yu2016negative}, both the pump pulse spectrum, spanning 1010~nm to 1140~nm (see Figure~\ref{fig:GMNA_FROG}(a)), and the near-infrared and visible spectral region extending down to 380~nm, are located within the fundamental guidance band, with losses below $\qty{0.3}{\dB\per\m}$. This increases the conversion efficiency from the pump pulse to the RDW, as it avoids crossing a high-loss resonance region, which disturbs the RDW generation process~\cite{Tani:18}. The fiber length was chosen to be slightly longer than the soliton fission length for the pump pulses and gas pressures used. The antiresonant fiber is sealed between two gas cells, with optical access provided by two 3~mm thick uncoated fused-silica windows. At the entrance of the hollow-core fiber, 350~nJ of pulse energy is available. We coupled into the fundamental mode and achieved up to 87\% coupling efficiency.

Figure~\ref{fig:GMNA_FROG}(a) presents the experimental amplifier output spectrum with increasing output pulse energy. The maximum pulse energy obtained before observing spectral instability was approximately 520~nJ, with a spectral bandwidth of around 122~nm at $1/\mathrm{e}^2$ of the peak. As the pulse energy increases, the pulse spectral moment shifts to longer wavelengths due to red shifting gain in the amplifier, which is consistent with the GMNA dynamics~\cite{Sidorenko:19}. We hypothesize that the spectral spike around 1030~nm is due to excess amplified spontaneous emission after amplification.

\begin{figure}
\centering
\includegraphics{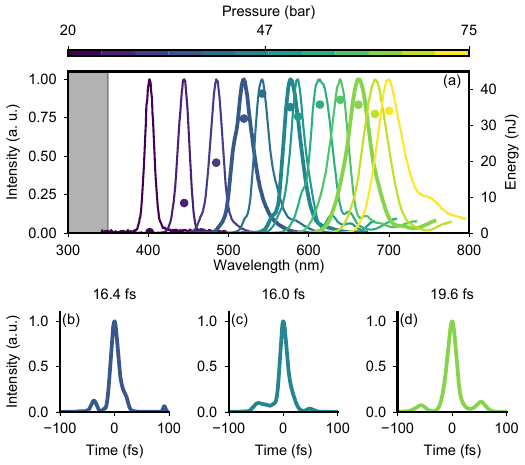}
\caption{\label{fig:tunability} (a) Tunable RDW generation (left axis) in a \qty{36}{\um} core diameter antiresonant fiber as the argon filling pressure is varied from 20 to 75 bar. The scattered dots (right axis) represent the energy of the corresponding RDWs at the fiber output. The gray-shaded area indicates the fiber's first high-loss resonance band. (b)-(d) Retrieved temporal pulse profiles corresponding to the bold spectra in (a). These correspond to (b) 520~nm (35~bar), (c) 576~nm (45~bar), and (d) 660~nm (65~bar). The spectra presented here are collected using an integrated sphere connected to a fiber-coupled CCD spectrometer (AvaSpec-ULS2048CL-EVO). The spectrometer covers the spectral range 200~nm to 1200~nm. The whole system is calibrated on an absolute scale with NIST traceable lamps.}
\end{figure}

To characterize the temporal profile of the GMNA output pulse, we employed a self-built all-reflection second-harmonic generation frequency-resolved optical gating (SHG-FROG). Figure~\ref{fig:GMNA_FROG}(b) shows the retrieved pulse temporal profile with a duration of 38~fs at FWHM. With more than 75\% of the pulse energy concentrated in the main pulse, the peak power reaches approximately \qty{10}{\MW}. The retrieved spectrum is presented in Figure~\ref{fig:GMNA_FROG}(a) as a dashed line showing a good agreement with the measured spectra. This pulse duration is comparable to that achievable with Ti:sapphire amplifiers and is sufficiently short to drive RDW generation directly.

Figure~\ref{fig:tunability} shows the generated RDW spectrum from around 400~nm to 700~nm, obtained by varying the argon filling pressure from 20 to 75~bar to tune the phase-matching condition for the RDW~\cite{Mak:13}. Further tuning beyond 700~nm was also achieved at higher pressures. However, for higher gas pressure, the RDW does not cleanly separate from the pump pulse, instead it forms part of a supercontinuum. The RDW energy is represented by the scattered points. These energies were extracted from the calibrated spectra and validated using a power meter and long-pass filter. The RDW energy peaks at around 540~nm with 39~nJ, resulting in a conversion efficiency of approximately 13\% from the coupled energy. This high conversion efficiency is attributed to the absence of a high-loss resonance band in the fiber's transmission spectrum between the pump pulse wavelength and the RDW wavelength. The gradual drop in RDW energy below 500~nm is due to insufficient pump pulse energy (as opposed to interference from the high-loss resonance band). Shorter wavelength RDW generation occurs at lower gas pressure (and hence lower nonlinearity) and larger frequency shifts from the pump pulse. Both of these reduce the RDW conversion efficiency, as predicted theoretically~\cite{Biancalana2004}, and observed in many experiments~\cite{Mak:13,Travers2019}.

\begin{figure}[t!]
\centering
\includegraphics{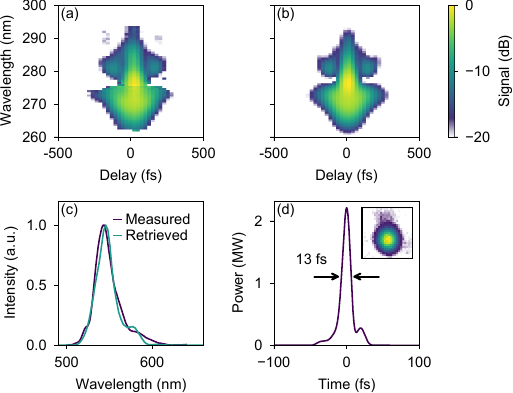}
\caption{\label{fig:FROG} (a) The measured and (b) the retrieved FROG traces of the RDW pulse generated at 40~bar. (c) The measured and FROG-retrieved spectrum of the RDW. (d) The retrieved temporal profile of the pulse. The inset in (d) shows the near-field beam profile of the RDW.}
\end{figure}

We conducted experiments with the system running for tens of hours without observing any degradation in either amplifier or hollow-core fiber performance. To further assess system stability, we characterized both the short-term and long-term fluctuations. A filtered RDW pulse train recorded with a 150 MHz photodiode confirmed stable pulse emission with 0.8\% relative intensity noise, which is only slightly higher than the 0.7\% relative intensity noise of the GMNA output. Further optimization and detailed analysis of the noise properties of this system will be reported elsewhere.

At the generation point, the RDW pulse is close to its transform-limited duration~\cite{Ermolov:16}. However, due to further propagation inside the fiber, output window, collimating lens, and air path, the RDW pulse temporally broadens. To compress the RDW, we first spectrally filter the RDW from the rest of the spectrum using an 850~nm long-pass filter. We then send the RDW to a set of chirped mirrors (PC70, Ultrafast Innovations) to compensate for the positive dispersion acquired after the generation point. We characterized the compressed RDW pulses using an all-reflection SHG-FROG. The crystal used is a type-I \qty{10}{\um}-thick $\beta$-barium borate (BBO) crystal. Figure~\ref{fig:FROG}(a) and (b) show the measured and retrieved FROG traces respectively. A good agreement is obtained between the traces, with 0.8\% error between them for a $512 \times 40$ grid size. Figure~\ref{fig:FROG}(d) shows the retrieved temporal profile for the pulse with a duration of 13~fs (FWHM) and more than 81\% of the energy located within the main peak. This corresponds to an estimated peak power of 2.2~MW. Figure~\ref{fig:FROG}(d) inset shows the near-field beam profile for the RDW characterized in the same figure, demonstrating a fundamental mode profile. Although much shorter RDW pulses have been obtained previously in antiresonant fibers~\cite{Ermolov:16, Brahms:19}, the limiting factor here is the RDW bandwidth, which increases as the RDW is shifted further from the pump pulse frequency. For example, in Ref.~\cite{Ermolov:16}, the RDW frequency shift from the pump pulse was around 750~THz and the RDW pulse duration was around 4~fs, while for Figure~\ref{fig:FROG} the frequency shift was only around 200~THz, and the pulse duration 13~fs. We also characterized the pulse duration at 660~nm (65~bar), 576~nm (45~bar), and 520~nm (35~bar), shown in Figure~\ref{fig:tunability}(b)-(d). We measured a pulse duration less than 20~fs in all cases.

The presented results demonstrate superior performance compared to current state-of-the-art tunable sources in terms of optical performance (duration, tuning range, peak power) and simplicity (cost, compactness). While our setup currently includes several free-space optical components for experimental convenience, they can all be further miniaturized or fiber-integrated. In addition, further enhancements to the system are possible. The first two stages, namely the oscillator and the pre-amplifier, could be substituted with a single, more powerful, oscillator that possesses the desired repetition rate~\cite{Erkintalo:12}. Furthermore, the entire oscillator and GMNA could be replaced with a single Mamyshev oscillator~\cite{Liu:19}. This modification would simplify the system and enhance its compactness. The employment of a large-mode-area photonic crystal fiber in this study was intended to demonstrate the most energetic RDW we were able to achieve. It is feasible to achieve tunable pulses with reduced---but adequate---energy using a standard step-index gain fiber~\cite{Sabbah:24_2}. In contrast, by moderately increasing the pump pulse energy (with a more optimized GMNA), further RDW tuning towards the resonance, below \qty{400}{\nm}, would be feasible.

Our approach offers the potential for expanded tunability into the ultraviolet and infrared spectral domains. As we demonstrated in recent work (ref.~\cite{Sabbah:24}), deep-ultraviolet RDW generation is achievable with pump energies as low as approximately 20~nJ in the green spectral region, a level readily accessible via SHG of the laser system described here. Moreover, by filling the antiresonant fiber with a Raman active gas, soliton self-frequency redshifting can be achieved~\cite{Tani2022, Chen:22}. Consequently, by incorporating an SHG stage and utilizing an appropriate set of antiresonant fibers, a single device could generate tunable ultrashort pulses spanning from the deep ultraviolet to the infrared.

In this work, we demonstrated the generation of tunable sub-20~fs pulses from around 400~nm to beyond 700~nm using a fiber oscillator and gain-managed nonlinear amplifier combined with RDW emission in a gas-filled hollow-core fiber. We achieved pulse durations as short as 13~fs with up to 39~nJ of energy and peak powers as high as 2.2~MW. This compact and efficient setup, operating at 4.8~MHz, offers a compelling alternative to traditional ultrafast tunable light sources, providing comparable or superior performance in terms of pulse duration and tunability while significantly reducing complexity, cost, and footprint. This work highlights the potential of GMNA and RDW techniques to create high-performance, low-cost, ultrafast laser sources suitable for a wide range of applications beyond laboratory settings.

\begin{backmatter}
\bmsection{Funding} This work was funded by the United Kingdom's Engineering and Physical Sciences Research Council: Grant agreement EP/T020903/1. JCT is supported by a Chair in Emerging Technologies from the Royal Academy of Engineering and by the Institution of Engineering and Technology (IET) through the IET A F Harvey Engineering Research Prize.

\bmsection{Acknowledgments} The authors thank M. Gebhardt, C. Brahms, W. J. Wadsworth, J. Knight and R. Thomson for useful discussions.

\bmsection{Disclosures} The authors declare no conflicts of interest.

\bmsection{Data availability} Data may be obtained from the authors upon reasonable request.

\end{backmatter}

\bibliography{refs}

\begin{thebibliography}{10}
\newcommand{\enquote}[1]{``#1''}

\bibitem{Denk1990}
W.~Denk, J.~H. Strickler, and W.~W. Webb, {\protect\JournalTitle{Science}} \textbf{248}, 73 (1990).

\bibitem{Maiuri2020}
M.~Maiuri, M.~Garavelli, and G.~Cerullo, {\protect\JournalTitle{Journal of the American Chemical Society}} \textbf{142}, 3 (2020). PMID: 31800225.

\bibitem{Kotsina_2022}
N.~Kotsina, C.~Brahms, S.~Jackson, \emph{et~al.}, {\protect\JournalTitle{Chem. Sci.}} \textbf{13}, 9586 (2022).

\bibitem{Ell2001}
R.~Ell, U.~Morgner, F.~X. K\"{a}rtner, \emph{et~al.}, {\protect\JournalTitle{Opt. Lett.}} \textbf{26}, 373 (2001).

\bibitem{Cerullo2003}
G.~Cerullo and S.~De~Silvestri, {\protect\JournalTitle{Review of Scientific Instruments}} \textbf{74}, 1 (2003).

\bibitem{Deckert:23}
T.~Deckert, A.~Vanderhaegen, and D.~Brida, {\protect\JournalTitle{Opt. Lett.}} \textbf{48}, 4496 (2023).

\bibitem{Mevert:21}
R.~Mevert, Y.~Binhammer, C.~M. Dietrich, \emph{et~al.}, {\protect\JournalTitle{Photon. Res.}} \textbf{9}, 1715 (2021).

\bibitem{Haohua2013}
H.~Tu and S.~A. Boppart, {\protect\JournalTitle{Laser \& Photonics Reviews}} \textbf{7}, 628 (2013).

\bibitem{Joly:11}
N.~Y. Joly, J.~Nold, W.~Chang, \emph{et~al.}, {\protect\JournalTitle{Phys. Rev. Lett.}} \textbf{106}, 203901 (2011).

\bibitem{Mak:13}
K.~F. Mak, J.~C. Travers, P.~H\"{o}lzer, \emph{et~al.}, {\protect\JournalTitle{Opt. Express}} \textbf{21}, 10942 (2013).

\bibitem{Belli:15}
F.~Belli, A.~Abdolvand, W.~Chang, \emph{et~al.}, {\protect\JournalTitle{Optica}} \textbf{2}, 292 (2015).

\bibitem{Mak:15}
K.~F. Mak, M.~Seidel, O.~Pronin, \emph{et~al.}, {\protect\JournalTitle{Opt. Lett.}} \textbf{40}, 1238 (2015).

\bibitem{Kottig:17}
F.~K\"{o}ttig, F.~Tani, C.~M. Biersach, \emph{et~al.}, {\protect\JournalTitle{Optica}} \textbf{4}, 1272 (2017).

\bibitem{Travers2019}
J.~C. Travers, T.~F. Grigorova, C.~Brahms, and F.~Belli, {\protect\JournalTitle{Nature Photonics}} \textbf{13}, 547 (2019).

\bibitem{Erkintalo:2012}
M.~Erkintalo, Y.~Q. Xu, S.~G. Murdoch, \emph{et~al.}, {\protect\JournalTitle{Phys. Rev. Lett.}} \textbf{109}, 223904 (2012).

\bibitem{Brahms:19}
C.~Brahms, D.~R. Austin, F.~Tani, \emph{et~al.}, {\protect\JournalTitle{Opt. Lett.}} \textbf{44}, 731 (2019).

\bibitem{Reduzzi:23}
M.~Reduzzi, M.~Pini, L.~Mai, \emph{et~al.}, {\protect\JournalTitle{Opt. Express}} \textbf{31}, 26854 (2023).

\bibitem{dudley_supercontinuum_2006}
J.~M. Dudley, G.~Genty, and S.~Coen, {\protect\JournalTitle{Reviews of Modern Physics}} \textbf{78}, 1135 (2006).

\bibitem{Sabbah:24}
M.~Sabbah, K.~Harrington, L.~R. Murphy, \emph{et~al.}, {\protect\JournalTitle{Opt. Lett.}} \textbf{49}, 3090 (2024).

\bibitem{Liu:19}
W.~Liu, R.~Liao, J.~Zhao, \emph{et~al.}, {\protect\JournalTitle{Optica}} \textbf{6}, 194 (2019).

\bibitem{Repgen:19}
P.~Repgen, D.~Wandt, U.~Morgner, \emph{et~al.}, {\protect\JournalTitle{Opt. Lett.}} \textbf{44}, 5973 (2019).

\bibitem{Sidorenko:20}
P.~Sidorenko and F.~Wise, {\protect\JournalTitle{Opt. Lett.}} \textbf{45}, 4084 (2020).

\bibitem{Sidorenko:19}
P.~Sidorenko, W.~Fu, and F.~Wise, {\protect\JournalTitle{Optica}} \textbf{6}, 1328 (2019).

\bibitem{Fermann2000}
M.~E. Fermann, V.~I. Kruglov, B.~C. Thomsen, \emph{et~al.}, {\protect\JournalTitle{Phys. Rev. Lett.}} \textbf{84}, 6010 (2000).

\bibitem{tomaszewska2022}
D.~Tomaszewska-Rolla, R.~Lindberg, V.~Pasiskevicius, \emph{et~al.}, {\protect\JournalTitle{Scientific Reports}} \textbf{12}, 404 (2022).

\bibitem{REN2023}
B.~Ren, C.~Li, T.~Wang, \emph{et~al.}, {\protect\JournalTitle{Optics \& Laser Technology}} \textbf{160}, 109081 (2023).

\bibitem{Boulanger:23}
V.~Boulanger, M.~Olivier, F.~Tr\'{e}panier, \emph{et~al.}, {\protect\JournalTitle{Opt. Lett.}} \textbf{48}, 2700 (2023).

\bibitem{Jiang2016}
T.~Jiang, Y.~Cui, P.~Lu, \emph{et~al.}, {\protect\JournalTitle{IEEE Photonics Technology Letters}} \textbf{28}, 1786 (2016).

\bibitem{Mayer:20}
A.~S. Mayer, W.~Grosinger, J.~Fellinger, \emph{et~al.}, {\protect\JournalTitle{Opt. Express}} \textbf{28}, 18946 (2020).

\bibitem{Russell2003}
P.~Russell, {\protect\JournalTitle{Science}} \textbf{299}, 358 (2003).

\bibitem{PolettiPetrovichRichardson}
F.~Poletti, M.~N. Petrovich, and D.~J. Richardson, {\protect\JournalTitle{Nanophotonics}} \textbf{2}, 315 (2013).

\bibitem{Litchinitser:02}
N.~M. Litchinitser, A.~K. Abeeluck, C.~Headley, and B.~J. Eggleton, {\protect\JournalTitle{Opt. Lett.}} \textbf{27}, 1592 (2002).

\bibitem{yu2016negative}
F.~Yu and J.~Knight, {\protect\JournalTitle{IEEE Journal of Selected Topics in Quantum Electronics}} \textbf{22}, 1 (2016).

\bibitem{Tani:18}
F.~Tani, F.~K\"{o}ttig, D.~Novoa, \emph{et~al.}, {\protect\JournalTitle{Photon. Res.}} \textbf{6}, 84 (2018).

\bibitem{Biancalana2004}
F.~Biancalana, D.~V. Skryabin, and A.~V. Yulin, {\protect\JournalTitle{Phys. Rev. E}} \textbf{70}, 016615 (2004).

\bibitem{Ermolov:16}
A.~Ermolov, H.~Valtna-Lukner, J.~Travers, and P.~S. Russell, {\protect\JournalTitle{Opt. Lett.}} \textbf{41}, 5535 (2016).

\bibitem{Erkintalo:12}
M.~Erkintalo, C.~Aguergaray, A.~Runge, and N.~G.~R. Broderick, {\protect\JournalTitle{Opt. Express}} \textbf{20}, 22669 (2012).

\bibitem{Sabbah:24_2}
M.~Sabbah, R.~Mears, L.~Murphy, \emph{et~al.}, \enquote{Tuneable megawatt-scale sub-15 fs visible pulses via dispersive wave emission pumped by a gain managed fiber amplifier,} in \emph{Laser Congress 2024 (ASSL, LAC, LS\&C),}  (Optica Publishing Group, 2024), p. AW6A.4.

\bibitem{Tani2022}
F.~Tani, J.~Lampen, M.~Butryn, \emph{et~al.}, {\protect\JournalTitle{Phys. Rev. Appl.}} \textbf{18}, 064069 (2022).

\bibitem{Chen:22}
Y.-H. Chen, P.~Sidorenko, E.~Antonio-Lopez, \emph{et~al.}, {\protect\JournalTitle{Opt. Lett.}} \textbf{47}, 285 (2022).

\end{thebibliography}

\bibliographyfullrefs{refs}

\end{document}